# An invisibility cloak using silver nanowires

*Yangbo Xie[1, 3], Huanyang Chen[1,] \*, Yadong Xu[1], Lin Zhu[1, 2], Hongru Ma[2], and Jian-Wen Dong[3]*

[1] *School of Physical Science and Technology, Soochow University, Suzhou, Jiangsu 215006, China*

[2] *Institute of Theoretical Physics, Shanghai Jiao Tong University, Shanghai 200240, China*

[3] *State Key Laboratory of Optoelectronic Materials and Technologies, Sun Yat-sen (Zhongshan) University, 510275, Guangzhou, China*

**Abstract**

In this paper, we use the parameter retrieval method together with an analytical effective medium approach to design a well-performed invisible cloak, which is based on an empirical revised version of the reduced cloak. The designed cloak can be implemented by silver nanowires with elliptical cross-sections embedded in a polymethyl methacrylate host. This cloak is numerically proved to be robust for both the inner hidden object as well as incoming detecting waves, and is much simpler thus easier to manufacture when compared with the earlier proposed one [Nat. Photon. 1, 224 (2007)].

**1. Introduction**

Transformation optics [1, 2] has paved the way to control the electromagnetic fields and design devices with novel properties [3, 4]. The invisibility cloak [2] is one of the most intriguing applications. Due to the great difficulty of designing and implementing an ideal cloak, several important reduced versions [5, 6, 7] have been proposed with less challenging material parameters. The first experimental breakthrough [8] is the implementation of a two dimensional (2D) transverse electric (TE) reduced cloak [5] for microwaves in 2006. Due to the causality condition [9], it is impossible to have an ideal cloak that can work in a broadband frequency range. The carpet cloak [10] was later proposed by sacrificing the perfect cloaking effect for bandwidth. It was further fabricated for both microwaves [11, 12] and optical frequencies [13, 14, 15]. Although the carpet cloak has been realized for optical frequencies, it is still an imperative to achieve an omnidirectional optical cloak, regardless of its limited functionality within a narrow band of frequencies. In fact, a theoretical design for a 2D transverse magnetic (TM) reduced cloak [6] was proposed shortly after the first TE cloak implementation [8]. The designed TM cloak is said to be made of a composite medium with silver nanoneedles embedded in a glass host [6]. However, it would be still extremely difficult to fabricate such a complex structure. In this paper, we will first propose an empirical revised version of the 2D TM reduced cloak from numerical calculations in Sec. 2. Next, a simple introduction will be given to an effective medium approach for composite medium in Sec. 3. In Sec. 4, we will then bring forth a detailed design process of such a revised version of the TM cloak using composite medium with silver nanowires embedded in a polymethyl methacrylate (PMMA) host, which would be much simpler to fabricate than before. Lastly, conclusions of the paper will be presented in Sec. 5.

---

\* kenyon@ust.hk

## 2. An empirical revised version of the reduced cloak

The material parameters of an ideal cloak for TM waves are [5, 6],

$$\mu_z = \left(\frac{b}{b-a}\right)^2 \frac{r-a}{r}, \quad \varepsilon_\theta = \frac{r}{r-a}, \quad \varepsilon_r = \frac{r-a}{r}, \tag{1}$$

where $a$ and $b$ are the inner and outer radii of the cloak. By keeping the principal refractive indices unchanged, a reduced cloak was proposed with its parameters [6]:

$$\mu_z = 1, \quad \varepsilon_\theta = \left(\frac{b}{b-a}\right)^2, \quad \varepsilon_r = \left(\frac{b}{b-a}\right)^2 \left(\frac{r-a}{r}\right)^2. \tag{2}$$

Such a cloak has no magnetic response, and thereby can be implemented with composite medium of metal and dielectrics in principle.

In this paper, we introduce a scaling factor $\eta$ to $\varepsilon_\theta$ for an empirical revised version,

$$\mu_z = 1, \quad \varepsilon_\theta = \eta\left(\frac{b}{b-a}\right)^2, \quad \varepsilon_r = \left(\frac{b}{b-a}\right)^2 \left(\frac{r-a}{r}\right)^2. \tag{3}$$

By calculating the total scattering cross section σ for different $\eta$ numerically, we empirically find a certain $\eta$ that has a minimum σ, which denotes better cloaking effect. For example, we set $a$ = 150 nm, $b$ = 450 nm and the wavelength $\lambda$ = 451.7 nm (these parameters will be used to design a realizable cloak in the following sections). Suppose the inner core is a perfect magnetic conductor (PMC) cylinder.

Table 1 shows the normalized total scattering cross section σ (normalized to a bare PMC cylinder with radius $a$ = 150 nm) for different $\eta$. As illustrated, when $\varepsilon_\theta = 2.55$, the total scattering cross section σ has a minimum value, which is about one third of that of the original reduced TM cloak. $\varepsilon_\theta = 2.55$ will be used in our further design.

| $\eta$ | 0.9778 | 1 | 1.0222 | 1.0444 | 1.0667 | 1.0889 | 1.1111 |
|---|---|---|---|---|---|---|---|
| $\varepsilon_\theta$ | 2.2 | 2.25 | 2.3 | 2.35 | 2.4 | 2.45 | 2.5 |
| σ | 0.385 | 0.32 | 0.26 | 0.208 | 0.166 | 0.135 | 0.116 |
| $\eta$ | 1.1333 | 1.1556 | 1.1778 | 1.2 | 1.2222 | 1.2444 | |
| $\varepsilon_\theta$ | 2.55 | 2.6 | 2.65 | 2.7 | 2.75 | 2.8 | |
| σ | 0.109 | 0.114 | 0.131 | 0.16 | 0.201 | 0.252 | |

Table 1 Normalized total scattering cross section σ for different $\eta$ and $\varepsilon_\theta$.

In figure 1, we plot the scattering patterns of different cloak types interacting with an incident TM plane wave from left to right in the $x$-direction. Fig. 1(a) shows the scattering pattern of an ideal cloak. The wave front is perfectly preserved after the wave propagates through the cloak. Fig. 1(b) shows the scattering pattern of the reduced cloak. The wave front is a bit disturbed by the cloak. Fig. 1(c) shows the scattering pattern of the current empirical revised version of the reduced cloak (with $\varepsilon_\theta = 2.55$). The wave front is recovered to almost as perfect as that of the ideal cloak.

This phenomenon can be perceived intuitively as such that the larger value of $n_r$ ($\sqrt{\varepsilon_\theta \mu_z} = \sqrt{2.55}$) in Fig. 1(c) draws the energy flow and keeps it from spreading out as that in Fig. 1(b) (with $n_r = \sqrt{\varepsilon_\theta \mu_z} = \sqrt{2.25}$) when propagating through the cloak [16]. Further analytical and numerical exploration of the mechanism for this empirical formula is expected.

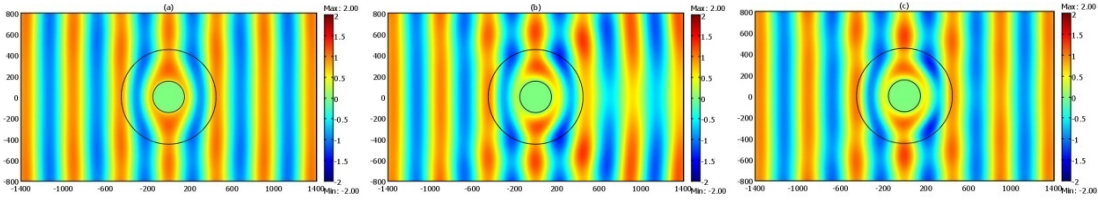

Fig. 1 The scattering patterns of (a) the ideal cloak, (b) the reduced cloak, and (c) the empirical revised version of the reduced cloak.

## 3. Effective Medium Approach

Before getting down to the detailed design of the above cloak, we first introduce an effective medium approach for composite medium, which is an initial method to obtain a rough design. Effective medium approaches (EMA) [17] are widely used to calculate the average electromagnetic response of a composite medium. Such a medium can be regarded as a macroscopic homogenous medium when the wavelength is much larger than the size of each unit cell. Maxwell–Garnett Theory (MGT) [18] is one of the most common approaches. Suppose we have a two-component composite medium with a permittivity of the host medium $\varepsilon_h$ and a permittivity of the inclusions $\varepsilon_i$. The inclusions are identical cylinders (both in shape and size) and are located in a two-dimensional square lattice inside the host medium with the lattice constant $l$. The primitive vectors are $\vec{a}_1 = l\,\hat{\imath}$ and $\vec{a}_2 = l\,\hat{\jmath}$, which are along the $x$-direction and $y$-direction. The cylinders are along the $z$-direction and have elliptical cross-sections with semiaxes $r_x$ and $r_y$ along the $x$ and $y$ coordinate axes respectively. Such a composite medium has an effective permittivity tensor $\overleftrightarrow{\varepsilon}_{eff} = diag\{\varepsilon_{eff,\ x},\ \varepsilon_{eff,\ y}\}$, which can be obtained from MGT with the depolarization or screening factor being taken into account [19],

$$\frac{\varepsilon_{eff,\ x} - \varepsilon_h}{\varepsilon_{eff,\ x} + u_x \times \varepsilon_h} = f \times \frac{\varepsilon_i - \varepsilon_h}{\varepsilon_i + u_x \times \varepsilon_h}\ , \qquad (4)$$

$$\frac{\varepsilon_{eff,\ y} - \varepsilon_h}{\varepsilon_{eff,\ y} + u_y \times \varepsilon_h} = f \times \frac{\varepsilon_i - \varepsilon_h}{\varepsilon_i + u_y \times \varepsilon_h}\ , \qquad (5)$$

where $u_x = \frac{r_x}{r_y}$, $u_y = \frac{r_y}{r_x}$ and $f = \frac{\pi r_x r_y}{l^2}$.

Given the aforementioned parameters of host medium and inclusions, the effective permittivity tensor $\overleftrightarrow{\varepsilon}_{eff}$ of the composite medium is only determined by $r_x$, $r_y$ and the filling ratio (or $l$). If the host medium is a dielectrics and the inclusions are metal cylinders with $-\mathcal{R}e(\varepsilon_i)$ being several times of $\varepsilon_h$ (i.e., the working frequency is slightly below the plasmonic frequency of the metal in discussion), $\varepsilon_{eff,\ x}$ will be near the range of (0, 1) while $\varepsilon_{eff,\ y}$ will be very close to $\varepsilon_h$ (suppose $r_x > r_y$). If we align the major axes ($x$-direction) with the radial direction (the $r$-direction) and the minor axes ($y$-direction) with the tangential direction (the $\theta$-direction), we can devise a rough framework of the above cloak with $\varepsilon_\theta$ close to $\varepsilon_h$ while $\varepsilon_r$ approximates the range of (0, 1). Therefore, the cloak can be implemented by embedding metal cylinders (nanowires) inside a dielectrics host medium, bringing in a design that could be fabricated more easily than before. This analytical tool might not be strictly precise, yet it can be used as estimation for parameters for further designing process in the next section.

## 4. Designing process and full-wave simulations

Apart from the EMA, parameter retrieval method (PRM) is another approach that can acquire the effective electromagnetic parameters much more accurately. PRM uses the numerically obtained reflection and transmission coefficients to extract the effective parameters [20, 21]. Based on the rough design from EMA, we can do the trial-and-error test to arrive at a precise design of the above cloak.

We set the inner and outer radii of the cloak to be 150nm and 450nm. The working wavelength is $\lambda$ = 451.7 nm (or 2.75eV, frequency in energy units). We will use PMMA as the host medium and silver nanowires with elliptical cross-sections as the inclusions. The permittivity of silver at this frequency [22] is -7.05805+0.21256*i, while that of PMMA [23] is 2.25.

We first break up the cloaking domain into five layers of equal thickness (60 nm). Each layer is divided into cells each measuring 60 nm in the middle. Each cell would have "fanlike" shape. The elliptic silver nanowires are placed in the center of each cell so that the major axes are all aligned with the radial direction (the $r$-direction). For each layer, the inclusions have different sizes of elliptical cross-sections. We use $r_r$ and $r_\theta$ to denote their semiaxes (which are related to $r_x$ and $r_y$ in Sec. 3). Here $l = 60\ nm$. After a very tedious trial-and-error test [24], we attain a precise design for the above cloak. The detailed sizes of the elliptic silver nanowires in each layer are shown in Table 2. The correspondingly retrieved electromagnetic parameters ($\varepsilon_r$ and $\varepsilon_\theta$) in each layer are also shown, which follows Eq. (3) considerably well (the absorption for this design is small).

| N | $r_r$ (nm) | $r_\theta$ (nm) | $\varepsilon_r$ | $\varepsilon_\theta$ |
|---|---|---|---|---|
| 1 | 20.3 | 5.1 | 0.151832+0.116957i | 2.653850+0.004352i |
| 2 | 19.4 | 4.7 | 0.369619+0.106619i | 2.598597+0.003679i |
| 3 | 18.4 | 4.4 | 0.506563+0.103219i | 2.556160+0.003193i |
| 4 | 19.4 | 4.1 | 0.817682+0.070841i | 2.539823+0.002916i |
| 5 | 19.1 | 3.9 | 0.931635+0.064073i | 2.518052+0.002662i |

Table 2 Parameters for each layer of the cloak

Now we will test the functionality of our designed cloak using the full-wave simulations. Firstly, as a reference, we plot the scattering pattern of a bare PMC cylinder with radius $a$ = 150 nm interacting with an incident TM plane wave in Fig. 2(a). Fig. 2(b) is the scattering pattern of the designed cloak with composite medium of silver nanowires and PMMA host. The wave front is preserved very well after the wave passes through the cloak except for a small shadow that comes from the absorption. To test the validity of the PRM and EMA, we plot the scattering pattern of a five-layer cloak in Fig. 2 (c) using the retrieved parameters ($\varepsilon_r$ and $\varepsilon_\theta$) in Table 2. The scattering pattern is visually the same as that in Fig. 2(b).

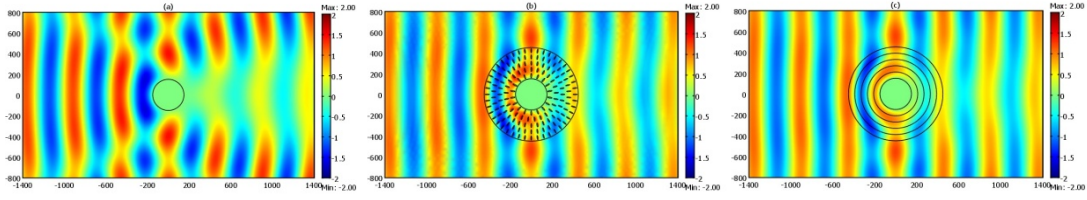

Fig. 2 The scattering patterns of (a) a bare PMC cylinder, (b) the designed cloak with silver nanowires embedded in a PMMA host, and (c) a five-layer cloak with the retrieved parameters in Table 2.

As a PMC cylinder here is just a model to verify the cloaking effect, it would be reasonable to test whether a dielectric core can be cloaked as well considering its availability. Fig. 3 (a) is the scattering pattern of the designed cloak by replacing the PMC cylinder with a PMMA cylinder. Fig. 3 (b) is the scattering pattern of the related five-layer cloak for comparison and validity testing. The scattering patterns tell that the cloak works very well for a dielectric core. For a metal core, we find that the cloak described by Eq. (3) works well (not shown here). By breaking up the cloaking domain into more than 10 layers instead of five, it will also bring in a good cloaking effect.

The designed cloak also works neatly for other kinds of wave sources. For example, we change the above TM plane wave into a cylindrical wave (excited by setting a constant magnetic field in a small circular boundary). Fig. 4 (a) shows the total field pattern for the designed cloak interacting with the cylindrical wave. Fig. 4 (b) shows the field pattern for the related five-layer cloak for validity test. The field patterns show that the cylindrical wave fronts are also well kept after passing through the cloak with little perturbation. Therefore, the above designed cloak causes little scattering and low loss, and is robust for different cores and incoming wave sources. The elliptical cross-sections of the nanowires in our design can also be approximated using other shapes, such as rectangular ones. The PRM is a useful tool to estimate the dimensions of the desired cross-sections. Furthermore, one can use other kinds of materials to design cloaks at other frequencies, for instance, using H-fractal metamaterials for microwave frequencies [25] and polaritonic materials for infrared (IR) frequencies [26].

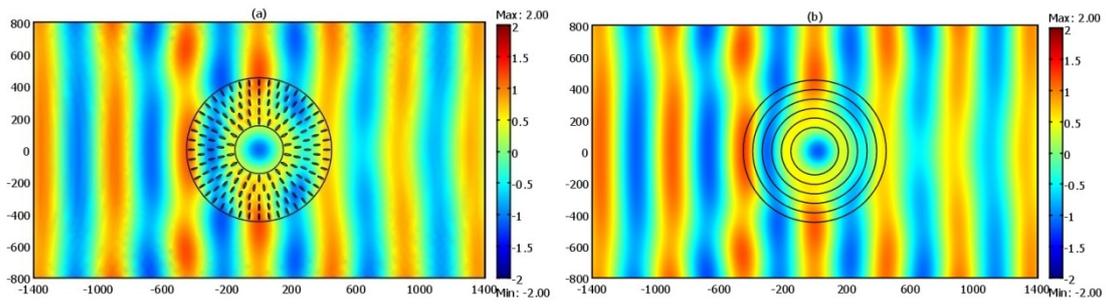

Fig. 3 The total filed patterns of (a) the designed cloak and (b) the five-layer cloak interacting with an incident TM plane wave. The inner core is a PMMA cylinder.

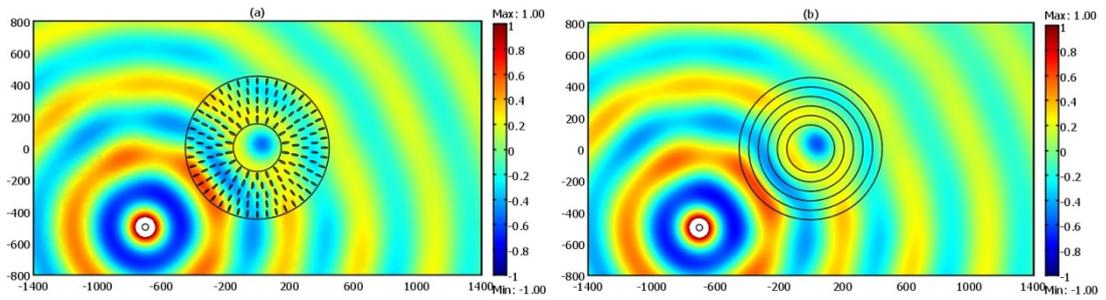

Fig. 4 The total filed patterns of (a) the designed cloak and (b) the five-layer cloak interacting with a cylindrical wave. The inner core is a PMMA cylinder.

## 5. Conclusion

In conclusion, the empirical revised version of the reduced cloak is tested to be more effective than the original reduced one. On this base, we propose an implementable design with an impressive cloaking effect. Its structure can be manufactured in a much easier way, and is likely to push forward the realization of practical invisible cloaking devices in the near future.

## Acknowledgements

This work was supported by the National Natural Science Foundation of China under Grant No. 11004147 and the Natural Science Foundation of Jiangsu Province under Grant No. BK2010211. Y.B.X and J.W.D. are supported by the National Natural Science Foundation of China under Grant No. 10804131 and the Guangdong Natural Science Foundation. L.Z and H.R.M. are supported by the National Natural Science Foundation of China under Grant No. 10874111.